\documentclass[%
 reprint,
superscriptaddress,
 amsmath,amssymb,amsfonts,
 aps,
prl,
 floatfix,
 longbibliography,nobibnotes
]{revtex4-2}

\usepackage[utf8]{inputenc}
\usepackage{comment} 
\usepackage{graphicx} %\graphicspath{{}}
\usepackage{xcolor}

\usepackage{bm}
\newcommand{\vect}[1]{\bm{#1}}

\usepackage{mathtools}

\DeclarePairedDelimiter\ket{\lvert}{\rangle}

\usepackage[colorlinks=true,linkcolor=blue,urlcolor=cyan,citecolor=blue]{hyperref}

\usepackage[final]{microtype}

\begin{document}

%\title{Gap protected scalable two-mode squeezing in power-law interacting 3-level spin systems}
\title{Gap-Protected Heisenberg-Limited Squeezing with Power-Law Interacting Multilevel Spins}
\author{Sakshi Bahamnia}
\affiliation{Homer L. Dodge Department of Physics and Astronomy, The University of Oklahoma, Norman, Oklahoma 73019, USA}
\affiliation{Center for Quantum Research and Technology, The University of Oklahoma, Norman, Oklahoma 73019, USA}
\author{Sagar Chaudhary}
\affiliation{Department of Physics, Oklahoma State University, Stillwater, Oklahoma 74078, USA}

\author{Arman Duha}
\affiliation{Department of Physics, Oklahoma State University, Stillwater, Oklahoma 74078, USA}
\author{Thomas Bilitewski}
\email{thomas.bilitewski@okstate.edu}
\affiliation{Department of Physics, Oklahoma State University, Stillwater, Oklahoma 74078, USA} 

\date{\today}

\begin{abstract}
We demonstrate a general mechanism to generate scalable, metrologically useful entanglement in the form of generalized spin squeezing in multi-level (qudit) spin systems with power-law interactions. 
We demonstrate this for spin-changing dynamics, extending the physics of spin-nematic squeezing known from spinor BECs to spatially extended lattice geometries.  In the absence of all-to-all connectivity, the dynamics generally suffers from leakage into non-collective finite-momentum modes, destroying Heisenberg scaling. We show that introducing $\mathfrak{su}(D)$ exchange interactions establishes an energy gap that protects the fully symmetric collective manifold, suppressing non-collective excitations, and achieving Heisenberg scaling $\xi^2\propto1/N$ in one to three dimensions for any power-law exponent. We provide an analytical understanding of the mechanism via the analysis of the Bogoliubov excitation spectrum, and demonstrate its effectiveness via numerical simulations of the quantum many-body dynamics.  Our findings more broadly apply to other spin-squeezing Hamiltonians, and directly generalize to larger internal spin dimensions. 
Our results provide a practical pathway towards scalable Heisenberg-limited qudit entanglement generation in experimental platforms such as arrays of (polar) molecules, magnetic atoms, and Rydberg atoms realizing multi-level spin systems.
\end{abstract}

\maketitle

{\it Introduction.---}%
Entanglement enables quantum technologies to perform beyond the capabilities of classical systems. In particular, it allows quantum sensors to surpass the standard quantum limit of uncorrelated particles. A particularly relevant and useful form of entanglement in the context of quantum-enhanced metrology \cite{Pezze2018,Degen2017,Giovannetti2011,Montenegro2025} is spin squeezing \cite{Wineland1992,Wineland1994,Kitagawa1993}, in which quantum correlations reduce measurement uncertainty below the classical limit. 

Scalable spin squeezing is well established theoretically and experimentally for systems with (effectively) all-to-all interactions in cavities, trapped ions, and Bose-Einstein condensates \cite{Wineland1992,Wineland1994,Kitagawa1993,Esteve2008,Riedel2010,Leroux2010,Lucke2011,Hamley2012,Hosten2016, Bohnet2016, PedrozoPenafiel2020,Greve_2022,Li_2023}. In Spinor BECs, spin-changing collisions create correlated pairs \cite{Law1998,Pu2000,Duan2000,Duan2002}, producing spin-nematic two-mode squeezing \cite{Hamley2012}, and twin-atom states \cite{Lucke2011,Gross2011,Bookjans2011,Luo2017}, and enabling quantum-enhanced clocks and interferometers \cite{Kruse2016,Linnemann2016,Mao2023}. 

Most programmable quantum simulators, however, feature power-law or short-range interactions \cite{Defenu2023,Kaufman2021}, including Rydberg atoms \cite{Saffman2010,Browaeys2020}, polar molecules \cite{Bohn2017,Yan2013,Christakis2023,Bao2023,Holland2023,Carroll2025,Wellnitz2024}, magnetic atoms \cite{Chomaz2023,Lepoutre2019,Patscheider2020}, and trapped ions \cite{Monroe2021}. Scalable squeezing has nonetheless been predicted \cite{Rey2008,Perlin2020,Bilitewski2021,Comparin2022a,Comparin2022b,Comparin2022c,Roscilde2023,Block2024,Roscilde2024}, and observed in Rydberg arrays \cite{Bornet2023,Eckner2023,Hines2023}, ion chains \cite{Franke2023}, lattice systems with contact interactions \cite{Lee2025}, magnetic atoms \cite{Douglas2025}, NV ensembles \cite{Wu2025}, and molecules \cite{HollandSq2026}. 

Scalable squeezing can be understood as arising from an energy gap separating the collective (Dicke) manifold from non-collective excitations \cite{Rey2008,Norcia2018,Perlin2020,Luo2025,Miller2024,Duha2024,Duha2025,Duha2026,Koyluoglu2025}, or, alternatively, from continuous-symmetry breaking and easy-plane order\cite{Comparin2022a,Comparin2022b,Comparin2022c,Roscilde2023,Block2024,Roscilde2024}.  Gap protection also underlies recent approaches to Heisenberg scaling with local interactions, via pair creation \cite{BilitewskiRey2023,Duha2024,Duha2025,Duha2026} and protected two-axis countertwisting \cite{Koyluoglu2025,Luo2025,Miller2024}.

Many of these platforms natively host $D>2$ internal levels (qudits), e.g., molecular rotational states~\cite{Bohn2017,Gorshkov2011}, Rydberg ladders~\cite{Ravets2014,Mogerle2025,Qiao2025,Zhang2024PRA,GonzalezCuadra2022}, large-spin magnetic atoms~\cite{Lepoutre2019,Patscheider2020,Chalopin2018}, and SU($n$)-symmetric alkaline-earth atoms~\cite{Gorshkov2010,Cazalilla2014}, whose larger operator algebra offers nematic (quadrupolar) quadratures absent for qubits~\cite{Hamley2012,Yukawa2013}. Multi-level squeezing has been explored for large-spin dipoles~\cite{Trifa2024}, cavity-coupled atoms~\cite{Sundar2023,Sundar2024}, SU($n$) fermions~\cite{Perlin2022}, and Rydberg arrays~\cite{Bahamnia2026}, but mechanisms for Heisenberg-limited scaling with finite-range interactions beyond spin-1/2 remain largely unexplored.

In this work, we identify a general gap-protection mechanism for scalable spin-squeezing generation in qudit systems, illustrated for the simplest case of a three-level (spin-1) system, and analyze under which conditions it enables Heisenberg-limited scaling of squeezing with finite-range interactions. We demonstrate the mechanism for two-mode squeezing generated by pair creation in power-law interacting spin-1 arrays, bringing spinor-BEC spin-nematic squeezing, with its Heisenberg-limited scaling, to programmable lattices of multi-level power-law interacting spin systems. Without gap protection, we find a dynamical transition from Heisenberg-limited to non-scalable squeezing as the interaction range decreases. With a local $\mathfrak{su}(3)$ gap protection term we show that Heisenberg scaling can be restored for any dimension $d$ and power-law exponent $\alpha$. We elucidate the origin of the gap protection via a Bogoliubov analysis of the excitation spectrum, providing an analytical understanding of both the general gap protection mechanism and the dynamical transition.

{\it Model.---}%
%%%%
\begin{figure}[tb]
\includegraphics[width=\columnwidth]{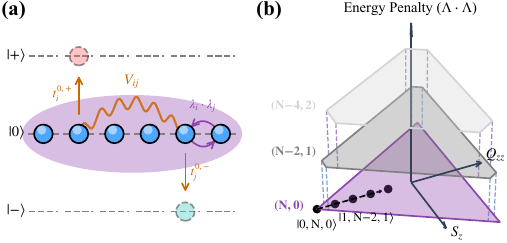}
\caption{Model and gap protection. (a) Power-law interacting spin-1 particles undergo pair creation $\ket{0,0}\to\ket{+1,-1}$, which generates entanglement; the local $\mathfrak{su}(3)$ exchange $V_{ij}\,\vect{\hat\lambda}_i\cdot\vect{\hat\lambda}_j$ protects the collective dynamics. (b) Hilbert-space sectors labeled by SU(3) irreducible representations $(p,q)$ and separated by gaps of the quadratic Casimir $\vect{\hat\Lambda}\cdot\vect{\hat\Lambda}$. In the collective sector $(N,0)$, states are labeled by occupations $\ket{N_{-1},N_0,N_{+1}}$; arrows indicate the states coupled by pair creation.\label{fig:fig1}}
\end{figure}
%%%%%%
We consider a $d$-dimensional array of $N$ three-level (spin-1) particles with sublevels $\ket{-1}$, $\ket{0}$, $\ket{+1}$, initialized in the polar product state $\vert \psi_0 \rangle =\vert 0 \rangle^{\otimes N}$. The particles interact via spin-changing pair processes
\begin{equation}
    H = \sum_{i \neq j} V_{ij} \, \left( \hat{t}^{0,+1}_i \hat{t}^{0,-1}_j + \mathrm{H.c.}\right),
    \label{eq:H_T}
\end{equation}
in terms of transition operators $\hat{t}_i^{m,m'} = \hat{b}_{i,m}^{\dagger} \hat{b}_{i,m'}$, where $\hat{b}_{i,m}^{\dagger}$ ($\hat{b}_{i,m}$) creates (annihilates) a particle at site $i$ in state $m\in\{-1,0,+1\}$, subject to $\sum_m \hat{b}^{\dagger}_{i,m}\hat{b}_{i,m}=1$. The interactions $V_{ij} = V\,(a/|\vect{r}_i - \vect{r}_j|)^{\alpha}$, with lattice spacing $a$ and coupling constant $V>0$, decay as a power law with exponent $\alpha$ with the distance between particles at positions $\vect{r}_i$, $\vect{r}_j$. %
Each term in Eq.~\eqref{eq:H_T} describes a coherent two-site pair process $\vert 0\rangle_i\vert 0\rangle_j \leftrightarrow \vert{+1}\rangle_i\vert{-1}\rangle_j$ converting a pair of $m=0$ particles into a correlated $(+1,-1)$ pair on sites~$i$ and~$j$ (as illustrated in Fig.~\ref{fig:fig1}(a)). %
%maybe some section on experimental realisations
Such processes arise, e.g., from spin-changing collisions in spinor gases~\cite{Law1998}, at F\"orster resonances of Rydberg atoms~\cite{Ravets2014}, and between rotational states of polar molecules~\cite{Gorshkov2011}. We provide explicit details in App.~\ref{app:exp_realisation}.

{\it Collective Limit.---} In the infinite-range limit ($V_{ij} = V$), the collective Hamiltonian decomposes into two independent parametric amplifiers operating on macroscopic spin-nematic quadratures $\mathcal{S}_\pm^{A/B}$ with $\mathcal{S}_\pm^A = \frac{1}{\sqrt{2N}}(S_x \pm Q_{yz})$, $\mathcal{S}_\pm^B = \frac{1}{\sqrt{2N}}(Q_{xz} \mp S_y)$ (App.~\ref{app:quadratures}). In the linearized regime, these modes exhibit exponential squeezing and anti-squeezing with variances $\text{Var}[\mathcal{S}_\pm] = e^{\pm 2\kappa t}$ at a collective rate $\kappa = NV$. Before depletion sets in at $e^{2\kappa t} \sim N$, the squeezed quadratures ($\mathcal{S}_-$) yield Heisenberg-limited metrological gain, $\xi^2 \sim 1/N$.

{\it Symmetric States and Gap Protection.---}%
The fully permutationally symmetric states of $N$ spin-1 particles can be labeled by occupation numbers $\ket{N_{-1},N_0,N_{+1}}$ and span the $(p,q)=(N,0)$ irreducible representation of SU(3), of dimension $(N+1)(N+2)/2$; all other sectors $(p,q)$ appearing in the $N$-fold tensor product carry non-collective excitations (Schur-Weyl duality \cite{Bacon2006,Jacak2021,RevModPhys.84.711}). Since the pair-creation dynamics conserves $S_z=N_{+1}-N_{-1}$, the collective dynamics starting from $\ket{0,N,0}$ is confined to the states $\ket{k,N-2k,k}$ as illustrated in Fig.~\ref{fig:fig1}(b).

The quadratic Casimir $\vect{\hat{\Lambda}}\cdot\vect{\hat{\Lambda}}=\sum_{a=1}^8\hat{\Lambda}_a^2$, with $\hat{\Lambda}_a=\sum_i\hat{\lambda}_a^{(i)}$, takes the value $\tfrac{4}{3}(p^2+q^2+pq+3p+3q)$ on the sector $(p,q)$ and is maximal on $(N,0)$; the gap to the nearest sector $(N-2,1)$ is $4N$. A term proportional to $\vect{\hat{\Lambda}}\cdot\vect{\hat{\Lambda}}$ therefore energetically isolates the fully symmetric manifold from all non-collective sectors (as illustrated in Fig.~\ref{fig:fig1}(b)).%, which is the basis of the gap-protection mechanism introduced below.

Motivated by this collective construction, we introduce a local $\mathfrak{su}(3)$ exchange with the same power-law couplings,
\begin{equation}
    H_{\text{prot}}=\frac{\chi}{4}\sum_{i \neq j}V_{ij}\sum_{a=1}^8\hat\lambda_a^{(i)}\hat\lambda_a^{(j)}=\frac{\chi}{2}\sum_{i\neq j}V_{ij}\Big(\hat P_{ij}-\frac13\Big),
    \label{eq:H_prot}
\end{equation}
where we used $\vect{\hat\lambda}_i\cdot\vect{\hat\lambda}_j=2\hat P_{ij}-2/3$, with $\hat P_{ij}$ the permutation operator swapping the states of particles $i$ and $j$. Since $\hat P_{ij}=1$ on all symmetric states, $H_{\rm prot}$ is constant on the entire collective manifold for \emph{arbitrary} $V_{ij}$, including disordered arrays, and commutes with all collective generators $\hat\Lambda_a$. Thus, it does not affect the collective dynamics, while penalizing non-collective excitations. For spin-1/2, $\hat P_{ij}=(1+\vect{\hat\sigma}_i\cdot\vect{\hat\sigma}_j)/2$, recovering the Heisenberg exchange that protects collective spin-1/2 dynamics~\cite{Rey2008,Perlin2020,Bilitewski2021,Duha2024,Koyluoglu2025}. For spin 1, by contrast, $\hat P_{ij}=\vect{\hat s}_i\cdot\vect{\hat s}_j+(\vect{\hat s}_i\cdot\vect{\hat s}_j)^2-1$, and we see that Heisenberg interactions alone are not constant on the symmetric manifold, and protecting qudit dynamics requires the full $\mathfrak{su}(3)$ exchange. The construction holds for any $D$, since $\vect{\hat\lambda}_i\cdot\vect{\hat\lambda}_j=2\hat P_{ij}-2/D$ for $\mathfrak{su}(D)$.

{\it Squeezing and numerical method.---}%
We quantify the metrological gain by the generalized squeezing parameter $\xi^2_{\rm opt}$, optimized over all linear $\mathfrak{su}(3)$ generators $\vect n\cdot\vect{\hat\Lambda}$ and read-out observables $\vect m\cdot\vect{\hat\Lambda}$ (see Appendix~\ref{app:squeezing}). %
$\xi^2_{\rm opt}<1$ certifies metrologically useful entanglement and $\xi^2_{\rm opt}\propto1/N$ corresponds to Heisenberg scaling. We denote by $\xi^2_{\min}$ its minimum over time and extract $\nu$ from $\xi^2_{\min}\propto N^{-\nu}$. %
We simulate the dynamics with the generalized discrete truncated Wigner approximation (GDTWA)~\cite{Schachenmayer2015,Zhu2019}, previously used for multi-level dipoles~\cite{Lepoutre2019,Patscheider2020,AzizAlaoui2024} and validated against exact methods for power-law spin-1/2 models~\cite{Muleady2023}. %
We provide benchmarks for the present model in Appendix~\ref{app:gdtwa}. %
We consider chains (1D), square (2D), and cubic (3D) lattices of linear size $L$ ($N=L^d$) with periodic boundary conditions, and set $\hbar=a=V=1$.

{\it Dynamical Squeezing Regimes.---} % no Gap protection terms
%%%%%%%%%%%%%%%%
\begin{figure}[tb]
\includegraphics[width=\columnwidth]{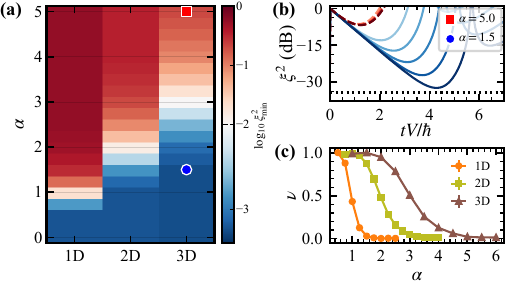}
\caption{Phases of squeezing dynamics without gap protection ($\chi=0$). (a) Minimal squeezing $\xi^2_{\rm min}$ (color scale) as a function of dimension $d$ and power-law exponent $\alpha$, showing the collective regime (blue, $\xi^2_{\rm min} \sim 1/N$) and the non-collective regime ($\xi^2_{\rm min} \sim N^{-\nu}$, $\nu<1$), for $L_{\mathrm{1D}}=10^4$, $L_{\mathrm{2D}}=100$, $L_{\mathrm{3D}}=22$. (b) Time evolution of $\xi^2_{\rm opt}$ in 3D for a characteristic $\alpha$ in the collective regime (blue, $\alpha = 1.5$) and in the non-collective regime (red, $\alpha = 5.0$), for $L_{\mathrm{3D}}=5,7,10,15,22$ from faint to dark. (c) Extracted scaling exponent $\nu$ of $\xi^2_{\rm min}\propto N^{-\nu}$ versus $\alpha$ for 1D, 2D and 3D (error bars smaller than symbol size).
\label{fig:fig2}}
\end{figure}
%%%%%%%%%%%%%%%%
% Time Dynamics of squeezing; comparison with bogoliubov/two mode squeezing prediction
We first consider the bare pair-creation Hamiltonian, Eq.~\eqref{eq:H_T}, and establish scalable, Heisenberg-limited spin-1 spin-nematic squeezing for sufficiently long-range power-law interactions.

Figure~\ref{fig:fig2}(a) shows $\xi^2_{\min}$ versus $d$ and $\alpha$, and we observe a transition from large squeezing and entanglement generation for long-range interactions (small $\alpha$) to weak, non-scalable squeezing for shorter range interactions (large $\alpha$). %
Figure~\ref{fig:fig2}(b) shows the time evolution of the squeezing parameter for two characteristic values of $\alpha$, one in the collective and one in the non-collective squeezing regime, in 3D for a range of system sizes. We observe the initial exponential squeezing $\xi^2_{\rm opt} \sim e^{-2V_0 t}$, predicted for the collective model, followed by saturation at a minimum that improves with $N$ in the collective regime ($\alpha=1.5$ in 3D) but not in the non-collective regime ($\alpha=5$). 
We analyze the system-size scaling of the best squeezing, $\xi^2_{\min}\propto N^{-\nu}$, in Fig.~\ref{fig:fig2}(c). In all dimensions, $\nu\simeq1$ for $\alpha\lesssim d/2$, followed by a continuous crossover centered at $\alpha=d$, where $\nu\approx1/2$, to non-scalable squeezing, $\nu\lesssim0.05$, for $\alpha\gtrsim3d/2$.

{\it Bogoliubov theory.---}
To analyze the spatial dynamics and finite-$\alpha$ effects, we linearize around the polar state using local Schwinger bosons $\hat{b}_{i,m}$. Since initially all particles occupy $m=0$, we replace $\hat{b}_{i,0} \to \sqrt{1- n_{i,1} - n_{i,-1}} \approx 1$, with $n_{i,m}=\hat{b}^{\dagger}_{i,m}\hat{b}_{i,m}$. This is valid as long as the excitation density remains small, $n_{i,\pm1}\ll 1$, i.e.\ during the exponential-growth stage that determines the collective squeezing. %

In momentum space the quadratic pair-production Hamiltonian becomes
%%%%%%%%%%
\begin{equation}
    H = \sum_k V_k \left( \hat{b}_{k,1}^\dagger \hat{b}_{-k,-1}^\dagger + \hat{b}_{k,1} \hat{b}_{-k,-1} \right),
    \label{eq:H_k_space}
\end{equation}
%%%%%%%%%%%%%%
where $\hat{b}_{i,m} = \frac{1}{\sqrt{N}} \sum_k e^{i \mathbf{k} \cdot \mathbf{r}_i} \hat{b}_{k,m}$, $V_k = \sum_{j \neq i} V_{ij} e^{i\mathbf{k} \cdot (\mathbf{r}_i - \mathbf{r}_j)}$, and $V_0 \equiv V_{k=0} = \sum_{j\neq i}V_{ij}$; in the all-to-all limit $V_0=(N-1)V\approx\kappa$. This directly demonstrates that the power-law interaction creates correlated pairs with opposite momenta. Each mode $(\mathbf{k},+1)$ is coupled to $( -\mathbf{k},-1)$ through pair creation and annihilation, generating two-mode squeezing between the paired modes.

The Heisenberg equations of motion yield the squeezing rate for each mode,
\begin{equation}
    \dot{\hat{b}}_{k,1} = -i V_k \hat{b}_{-k,-1}^\dagger, \qquad \dot{\hat{b}}_{-k,-1}^\dagger = i V_k \hat{b}_{k,1},
\end{equation}
i.e.\ every mode pair $(k,-k)$ is dynamically unstable with purely imaginary Bogoliubov frequency $\omega_k = \pm i V_k$ and growth exponent $\gamma_k = |V_k|$. For positive couplings $V_{ij}>0$ the collective $k=0$ mode grows fastest, $\gamma_0 = V_0 = \max_k |V_k|$, and the bosonic description reproduces the squeezed spin quadratures (App.~\ref{app:quadratures}).

%{\it Gap protection.---} 
Because $V_k\neq0$ for $\vect k\neq0$, the bare Hamiltonian also amplifies non-collective finite-$\vect k$ modes, which compete with the collective mode. At the squeezing time $t^*\simeq\ln(4N)/(2V_0)$ of the collective mode, mode $\vect k$ has $\sinh^2(V_kt^*)$ excitations, i.e., $\sim N^{V_k/V_0}$ for the most unstable modes. %
For $\alpha>d$, $V_k\to V_0$ as $k\to0$, thus, a continuum of soft modes grows almost as fast as the collective mode, and squeezing is expected not to be scalable. %
For $\alpha<d$, $V_k/V_0$ stays bounded below 1 for all $\vect k\neq0$, and the fraction of excitations in finite-$\vect k$ modes, $f\sim(t^*)^2\sum_{\vect k\neq0}V_k^2/N\sim(\ln N)^2\sum_jV_{ij}^2/V_0^2$ vanishes as $(\ln N)^2/N$ for $\alpha<d/2$, but only as $(\ln N)^2N^{-2(1-\alpha/d)}$ for $d/2<\alpha<d$. Linear theory thus predicts a transition at $\alpha=d$. For the accessible system sizes, $N\lesssim10^4$, the slow decay of $f$ below $\alpha=d$ and its slow growth just above may broaden this transition into the crossover of Fig.~\ref{fig:fig2}(c), where $\nu\simeq1$ for $\alpha\lesssim d/2$, $\nu\approx1/2$ at $\alpha=d$, and $\nu\approx0$ for $\alpha\gtrsim3d/2$.

Including the exchange interaction in Eq.~\eqref{eq:H_prot} shifts the energies of these modes while leaving the collective mode unchanged. In momentum space, this protection mechanism generates a mode-dependent energy gap
%%%%%%
\begin{equation}
    H_{\text{prot}} \approx  \chi \sum_k (V_k - V_0) \left( \hat{b}_{k,1}^\dagger \hat{b}_{k,1} + \hat{b}_{-k,-1}^\dagger \hat{b}_{-k,-1} \right)
\end{equation}
with $\hat{n}_{k,m}=\hat{b}_{k,m}^\dagger \hat{b}_{k,m}$. Defining the gap $\Delta_k = \chi(V_k - V_0)$, which vanishes identically at $k=0$ as a direct consequence of $H_{\rm prot}$ being constant on the symmetric manifold, the total momentum-space Hamiltonian is
\begin{equation}
    H_{\text{tot}} = \sum_k \left[ \Delta_k (\hat{n}_{k,1} + \hat{n}_{-k,-1}) + V_k (\hat{b}_{k,1}^\dagger \hat{b}_{-k,-1}^\dagger + \mathrm{H.c.}) \right].
\end{equation}
%%%%%
Diagonalizing $H_{\rm tot}$ yields the Bogoliubov dispersion $\omega_k=\pm\sqrt{\Delta_k^2-V_k^2}$. The collective mode ($k=0$) remains resonant, $\Delta_0=0$, and is amplified at the unchanged rate $\gamma_0=V_0$. A finite-$\vect k$ mode is stable (real $\omega_k$) if $|\Delta_k|>|V_k|$, and unstable with reduced rate $\gamma_k=\sqrt{V_k^2-\Delta_k^2}<\gamma_0$ otherwise. Requiring stability of all finite-momentum modes defines the critical protection strength $  \chi_{c,\mathrm{bog}} = \max_{k \neq 0} \frac{|V_k|}{V_0 - V_k}$. 
For $\chi > \chi_{c,\mathrm{bog}}$ only the $k=0$ mode is unstable, and we expect Heisenberg-limited scaling of the collective squeezing dynamics. 
\begin{figure}[tb]
%    \centering
    \includegraphics[width=\linewidth]{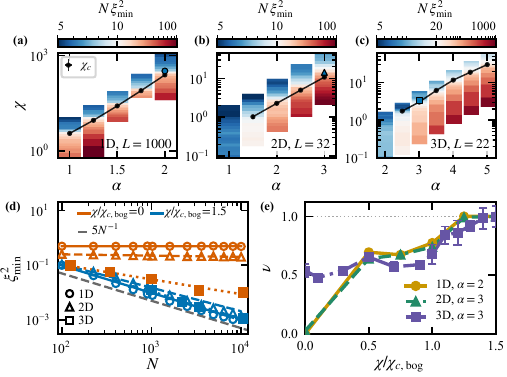}
    \caption{Gap-protected squeezing phase diagrams. (a-c) Minimal squeezing as a function of $\alpha$ and gap protection strength $\chi$ in 1D ($L=1000$) (a), 2D ($L=32$) (b) and 3D ($L=22$) (c). (d) Scaling of minimal squeezing versus system size $N$ for parameters below and above the transition with markers as indicated in the panels (a)-(c). (e) Scaling exponent of minimal squeezing $\xi^2_{\rm min} \propto  N^{-\nu}$ versus gap protection strength $\chi$ normalized by the Bogoliubov prediction $\chi_{c,\rm{bog}}$.
    \label{fig:fig3}}
\end{figure}

%%%%%%%%%%%%%

% Bogoliubov supplementary
The asymptotic system-size dependence of $\chi_{c,\mathrm{bog}}$ can be obtained analytically by evaluating $V_k$ at $k=2\pi/L$ yielding $\chi_c\to$ const.\ ($\alpha<d$), $\ln L$ ($\alpha=d$), $L^{\alpha-d}$ ($d<\alpha<d+2$), $L^2/\ln L$ ($\alpha=d+2$), $L^2$ ($\alpha>d+2$) (App.~\ref{app:critical-chi}).  As a direct consequence for $\alpha<d$ a fixed $\chi>\chi_{\infty}$ suffices at any $N$,  for $\alpha>d$, $\chi$ must grow with $L$. Specifically, for the relevant case of dipolar interactions ($\alpha=3$) $\chi$ needs to scale as $\ln L$ in 3D, $\propto L$ in 2D, $\propto L^2/\ln L$ in 1D. 

{\it Gap-protected Squeezing.---} %  Gap protection terms
% Phase Diagrams of model 1D, 2D, 3D; transition from scalable to non-scalable as function of alpha and gap protection strength
% Comparison with bogoliubov phase diagram
Next, we demonstrate that $\mathfrak{su}(3)$ gap protection indeed enables Heisenberg scaling of multi-level squeezing for any dimension $d$ and power-law exponent $\alpha$, as predicted by the Bogoliubov analysis.

Figures~\ref{fig:fig3}(a)-(c) show $\xi^2_{\mathrm{min}}$ as a function of the interaction exponent $\alpha$ and gap protection strength $\chi$ in 1D, 2D, and 3D. Across all dimensions, we observe a transition from a non-collective regime with small squeezing (red) to a fully collective regime with Heisenberg-limited scaling (blue) as the gap protection strength $\chi$ is increased. %
In Fig.~\ref{fig:fig3}(d) we demonstrate the system-size scaling of $\xi^2_{\mathrm{min}}$ in the collective ($\chi/\chi_{c,\mathrm{bog}}=1.5$, blue) and non-collective phase ($\chi/\chi_{c,\mathrm{bog}}=0$, orange) for all three spatial dimensions. For $\chi/\chi_{c,{\mathrm{bog}}}=1.5$ all three dimensions lie in the collective regime, and the squeezing exhibits Heisenberg scaling, $\xi_{\min}^2\sim  N^{-1}$. %
Figure~\ref{fig:fig3}(e) shows the extracted scaling exponent $\nu$, defined by $\xi_{\min}^2\sim N^{-\nu}$, as a function of $\chi/\chi_{c,{\mathrm{bog}}}$. At $\chi=0$, the scaling exponent is close to zero in 1D and 2D, indicating the absence of collective scaling, whereas 3D ($\alpha=d$) already exhibits a finite scaling exponent $0<\nu <1$. As $\chi/\chi_{c,{\mathrm{bog}}}$ increases, $\nu$ increases and reaches Heisenberg scaling $\nu=1$. For 1D and 2D, the transition occurs at $\chi/\chi_{c,{\mathrm{bog}}}\simeq1.25$, while in 3D at $\chi/\chi_{c,{\mathrm{bog}}}\simeq1.45$. 
%The system sizes used to extract the scaling exponents are $L=\{1000,1600,2500,3600,4900,6400,8100,10000\}$ in 1D, $L=\{32,40,50,60,70,80,90,100\}$ in 2D, and $L=\{10,13, 15, 22 \}$ in 3D.

We note that this also demonstrates that the Bogoliubov analysis quantitatively captures the transition and scaling of the required gap-protection strength (App.~\ref{app:critical-chi}). 
Thus, our results demonstrate that $\mathfrak{su}(3)$ gap protection enables Heisenberg-limited multi-level squeezing for any $\alpha$ and $d$, provided $\chi\gtrsim\chi_c(L)$.

{\it Site dilution.---} 
%%%%
\begin{figure}[tb]
    \centering
    \includegraphics[width=\linewidth]{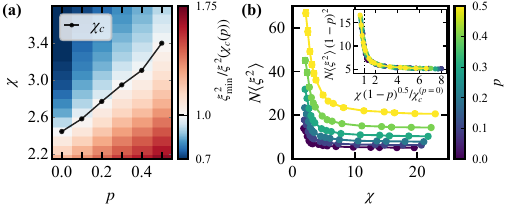}
    \caption{Robustness to site-dilution disorder. (a) Minimal squeezing $\xi^2_{\rm min}$, normalized by its value at $\chi_c(p)$, as a function of site-dilution probability $p$ and gap protection strength $\chi$ for dipolar interactions ($\alpha=3$) in 3D with $L=10$, $\chi_c(p)$ (triangles). (b) Minimal squeezing $N\xi^2_{\rm min}$ versus $\chi$ for different $p$ (colorbar). Inset: Scaling collapse of $N\xi^2_{\rm min}(1-p)^{2}$ versus $\chi(1-p)^{0.5}/\chi_c^{(p=0)}$.  $5000$ GDTWA trajectories, each with independently drawn dilution configuration.}% Errors smaller than symbol size.}
    \label{fig:fig4} 
\end{figure}
%%%%
Finally, we consider the robustness of the achievable squeezing to site-dilution disorder as relevant to tweezer and lattice experiments \cite{Christakis2023,HollandSq2026}. Generally, site dilution is expected to deteriorate squeezing as it breaks translational invariance, mixing collective and non-collective modes, although spin-1/2 squeezing can be robust to moderate disorder \cite{Kwasigroch2017,Begg2026,solfanelli2026robustspinsqueezingrandominteraction,KaplanLipkin2025}. 

Fig.~\ref{fig:fig4}(a) shows the resulting squeezing phase diagram in 3D  ($L=10$) for dipolar interactions ($\alpha=3$) as a function of dilution probability $p$ and gap protection strength $\chi$. Dilution drives the unprotected system into the non-collective phase, while increasing $\chi$ restores the collective phase at fixed $p$. 
In Fig.~\ref{fig:fig4}(b) we show the underlying raw data of the minimal squeezing versus $\chi$ for a range of $p$. Above a critical gap protection strength the minimal squeezing saturates at a value dependent on the site dilution $p$. The scaling collapse in the inset demonstrates that the plateau value scales as $1/(N (1-p)^2)$ and the critical gap protection strength as $\chi_c(p=0)/\sqrt{1-p}$.

{\it Conclusions.---} %
We have identified an $\mathfrak{su}(3)$ gap protection mechanism that enables Heisenberg-limited scaling of squeezing in locally interacting three-level spin models in any dimension $d$ and for any power-law exponent $\alpha$. We provide an understanding of this mechanism both in terms of the energy gaps of the different shells in Hilbert space, as well as an analytical description in terms of Bogoliubov analysis, which also predicts the correct scaling of the required gap protection strength with system size. The gap protection mechanism additionally protects against site dilution disorder. This enables Heisenberg-limited spin-nematic multi-level squeezing in programmable arrays of molecules, Rydberg atoms, and magnetic atoms realizing power-law or local interactions. 

The mechanism studied here straightforwardly generalizes to any multi-level system with $D$ internal states using the corresponding $\mathfrak{su}(D)$ Casimir operator, and any other collective squeezing dynamics, e.g., one- and two-axis twisting. Avenues for future work include protection by higher Casimir operators, interaction-based optimal read-out schemes, as well as multiparameter estimation in these multi-level qudit systems.

\begin{acknowledgments}
\noindent{\textit{Acknowledgments:}
TB acknowledges support from the National Science Foundation through NSF Grant No. PHY-2542371. Some of the computing for this project was performed at the OU Supercomputing Center for Education \& Research (OSCER) at the University of Oklahoma (OU). Some of the calculations were performed in the PETE system of the High Performance Computing Center at Oklahoma State University, NSF Grant No. OAC-1531128.
} 

\end{acknowledgments}

%\section*{Data Availability}
\noindent{\textit{Data Availability}}
The data that support the findings of this article are publicly available at \cite{data}.

% For references only cited in supplementary, add them here as nocite
%\nocite{}

% For arXiv, it is highly recommended to comment out the \bibliography line 
% and paste the contents of your .bbl file directly here, OR just upload 
% the .bbl file alongside this .tex file to arXiv.
%\input{main.bbl}
\bibliography{main}{}

\cleardoublepage
\appendix

% ==========================================
% SUPPLEMENTAL MATERIAL
% ==========================================

%%%%%%%%%% Prefix a "S" to all equations, figures, and reset the counter %%%%%%%%%%
\setcounter{equation}{0}
\setcounter{figure}{0}
\setcounter{table}{0}
\setcounter{section}{0}
\setcounter{secnumdepth}{2} % Re-enables section numbering disabled by the PRL documentclass

% Detach counters from section resets
\counterwithout*{equation}{section}
\counterwithout*{figure}{section}
\counterwithout*{table}{section}

\makeatletter
\renewcommand{\thesection}{\Alph{section}}
\renewcommand{\theequation}{S\arabic{equation}}
\renewcommand{\thefigure}{S\arabic{figure}}
\makeatother

\phantomsection
\section*{Supplemental Material}
The Supplemental Material contains details on the experimental realisation of the gap-protected model (App.~\ref{app:exp_realisation}), the operator definitions (App.~\ref{app:operators}), the mapping to phase-space quadratures (App.~\ref{app:quadratures}), the generalized squeezing parameter (App.~\ref{app:squeezing}), the GDTWA method and benchmarks (App.~\ref{app:gdtwa}), additional details on the Bogoliubov analysis (App.~\ref{app:critical-chi}), and extended data on the site-diluted case in 1D and 2D (App.~\ref{app:extended_data}).

\section{Experimental Realization}\label{app:exp_realisation}
The pair-creation model with gap protection, Eqs.~\eqref{eq:H_T} and~\eqref{eq:H_prot}, can be realized with polar molecules in optical lattices or tweezer arrays~\cite{Gorshkov2011,Bohn2017,Anderegg2019,Holland2023,Bao2023}, and in principle with Rydberg atoms at a Stark-tuned F\"orster resonance~\cite{Ravets2014,Ryabtsev2010}. A dc field makes pair creation between three (rotational) states resonant, and Floquet engineering via global microwave pulses transforms the native dipolar interactions into the desired gap-protected model. For specificity, we focus on polar molecules in the following.

\begin{figure}[t]
\centering
\includegraphics[width=\columnwidth]{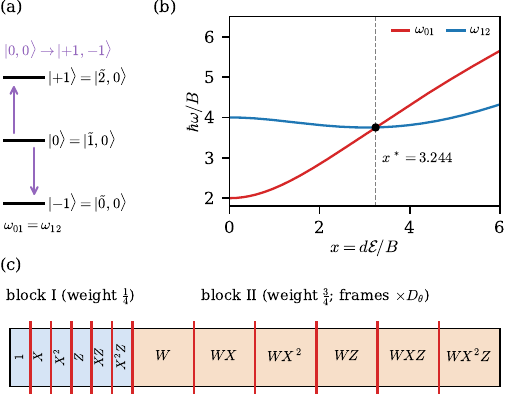}
\caption{(a) Rotational ladder and the resonant pair process. (b) Transition frequencies of the dressed rotor versus $x=d\mathcal E/B$. (c) Floquet pulse sequence that engineers the desired gap-protected interactions: toggling frames of block I ($X^sZ^k$) and block II ($WX^sZ^kD_\theta$), with box widths $\propto$ weights, separated by the global unitaries $G_f$ (red).}
\label{fig:si_mol}
\end{figure}

\emph{Rotational ladder.} A rigid rotor with rotational constant $B$ and dipole moment $d$ in a dc field $\mathcal E\hat z$, $H_{\rm rot}=B\hat{\mathbf N}^2-d_z\mathcal E$, has Stark-dressed eigenstates $\ket{\tilde N,M}$ (adiabatically connected to $\ket{N,M}$ at zero field) that depend only on $x=d\mathcal E/B$. We use
\begin{equation}
\ket{-1}=\ket{\tilde0,0},\qquad\ket{0}=\ket{\tilde1,0},\qquad\ket{+1}=\ket{\tilde2,0},
\end{equation}
with energies $E_m$, and label transition frequencies and dipole matrix elements by the rotor index $\tilde N=m+1\in\{0,1,2\}$, as illustrated in Fig.~\ref{fig:si_mol}(a). %
The energy spacings, $2B$ and $4B$ at zero field, depend on the electric field and coincide at $x^*=3.244$, $\hbar\omega_{01}=\hbar\omega_{12}\equiv\hbar\omega=3.752\,B$, as seen in Fig.~\ref{fig:si_mol}(b). At this resonance the energy defect $E_{-1}+E_{+1}-2E_0$ vanishes, and the pair-creation process $\ket{0,0}\to\ket{{+1},{-1}}$ is resonant. For common bialkali molecules the required field, $\mathcal E^*=x^*B/d$, is about 2--13\,kV/cm (Table~\ref{tab:si_species}). We note that the same crossing is used for resonant collisional shielding of KRb~\cite{Matsuda2020}.

\begin{table}[b]
\caption{Rotational constant and dipole moment of common bialkali molecules~\cite{Ni2008,Park2015,Guo2016,Molony2014,Cairncross2021}, and the resulting resonant field $\mathcal E^*=x^*B/d$.}
\label{tab:si_species}
\begin{ruledtabular}
\begin{tabular}{lccc}
 & $B/h$ (GHz) & $d$ (D) & $\mathcal E^*$ (kV/cm)\\
\hline
KRb  & 1.114 & 0.57  & 12.6\\
NaK  & 2.822 & 2.72  & 6.69\\
NaRb & 2.090 & 3.2   & 4.21\\
RbCs & 0.490 & 1.225 & 2.58\\
NaCs & 1.736 & 4.75  & 2.35\\
\end{tabular}
\end{ruledtabular}
\end{table}

\emph{Native and target interaction.} For molecules in a plane perpendicular to $\hat z$ ($\hat{\mathbf r}_{ij}\perp\hat z$), the $d_z^{(i)}d_z^{(j)}/(4\pi\epsilon_0r_{ij}^3)$ interaction is isotropic and acts within the $M=0$ manifold, whereas transverse dipole terms ($d_x,d_y$) change $M$ by $\pm1$ and are far off resonance. The corresponding matrix elements $\mu_{\tilde N\tilde N'}=\langle\tilde N,0|d_z|\tilde N',0\rangle/d$ at $x^*$ are
\begin{equation}
\mu=\begin{pmatrix}0.597&0.346&0.093\\0.346&-0.146&0.512\\0.093&0.512&-0.171\end{pmatrix},
\end{equation}
with respect to the state ordering $\ket{-1},\ket{0},\ket{+1}$.

Dropping off-resonant terms, we obtain $H_{\rm nat}=E_{\rm dd}\sum_{i<j}\mathcal V_{ij}h_{ij}$, with $E_{\rm dd}=d^2/4\pi\epsilon_0a^3$, $\mathcal V_{ij}=(a/r_{ij})^3$ and
\begin{align}
h_{ij}={}&\mu_{01}\mu_{12}\hat\Pi_{ij}+\mu_{01}^2\hat F^{-1,0}_{ij}+\mu_{12}^2\hat F^{0,+1}_{ij}\nonumber\\
&+\mu_{02}^2\hat F^{-1,+1}_{ij}+\hat\mu^{(i)}\hat\mu^{(j)}.
\label{eq:si_Hnat}
\end{align}
Here $\hat\Pi_{ij}=\hat\Pi^+_{ij}+{\rm \mathrm{H.c.}}$ with $\hat\Pi^+_{ij}=\hat t^{+1,0}_i\hat t^{-1,0}_j+\hat t^{-1,0}_i\hat t^{+1,0}_j$ creates and annihilates pairs, $\hat F^{m,m'}_{ij}=\hat t^{m,m'}_i\hat t^{m',m}_j+{\rm \mathrm{H.c.}}$ describes exchange $\ket{m,m'}\leftrightarrow\ket{m',m}$, and the last term is a density--density (Ising) coupling of the field-induced dipoles $\hat\mu^{(i)}=\mu_{00}\hat n_{i,-1}+\mu_{11}\hat n_{i,0}+\mu_{22}\hat n_{i,+1}$, with $\hat n_{i,m}=\hat t^{m,m}_i$ the occupation of level $m$. All coefficients are in units of $E_{\rm dd}$.

The gap-protected model, Eqs.~\eqref{eq:H_T} and~\eqref{eq:H_prot} with $\alpha=3$, can be rewritten as $V\sum_{i<j}\mathcal V_{ij}[\hat\Pi_{ij}+\chi(\hat P_{ij}-\tfrac13)]$. The swap $\hat P_{ij}=\sum_{m<m'}\hat F^{m,m'}_{ij}+\sum_m\hat n_{i,m}\hat n_{j,m}$ contains equal exchange between all three states and the density-density term $\sum_m\hat n_{i,m}\hat n_{j,m}$. The pulses must therefore keep the pair term, symmetrize the three exchange amplitudes, and turn the Ising term into a diagonal symmetric density-density interaction with the same strength as the exchange.

\emph{Floquet engineering.} %
Because $\mu_{12}/\mu_{01}=1.48\neq1$ at $x^*$, the two quadratures of a single resonant $\pi$-polarized microwave field at frequency $\omega$ allow any global single-molecule unitary $G_f$ to be implemented via composite pulses. We use these global unitaries to define toggling frames $U_f=G_f\cdots G_1$. For cycle times short compared with the interaction time $\hbar/(E_{\mathrm{dd}}\mathcal V_0)$, with $\mathcal V_0=\sum_{j\neq i}\mathcal V_{ij}$, the dynamics is generated by the average Hamiltonian~\cite{Choi2020} $H_{\mathrm{eff}}=E_{\mathrm{dd}}\sum_{i<j}\mathcal V_{ij}h^{\mathrm{eff}}_{ij}$ with
\begin{equation}
h^{\rm eff}_{ij}=\sum_fw_f\,U_f^{(ij)\dagger}h_{ij}\,U_f^{(ij)},\qquad U_f^{(ij)}=U_f^{(i)}U_f^{(j)},
\end{equation}
where $w_f$ is the fraction of the cycle spent in frame $f$.

For the pulses used below these averages have the form $(a\hat\Pi^+_{ij}+\mathrm{\mathrm{H.c.}})+c_P\hat P_{ij}+c_Q\sum_m\hat n_{i,m}\hat n_{j,m}+\mathrm{const}$. If $c_Q=0$, this yields the desired gap-protected model with $V=J'E_{\mathrm{dd}}$, $J'=|a|$, and $\chi=c_P/J'$ (up to a phase of $a$ that can be absorbed into $\ket{\pm1}$). Since $U_f^{(ij)}$ commutes with $\hat P_{ij}$, the trace $\mathrm{Tr}[(\hat P_{ij}-\tfrac13)h_{ij}]$ is unchanged by any global pulse sequence. It equals $2\sigma_\mu^2+6\overline{\mu^2}$ for $h_{ij}$, with the mean exchange strength $\overline{\mu^2}=\tfrac13(\mu_{01}^2+\mu_{12}^2+\mu_{02}^2)=0.130$ and the variance of the induced dipoles $\sigma_\mu^2=\tfrac16\sum_{\tilde N<\tilde N'}(\mu_{\tilde N\tilde N}-\mu_{\tilde N'\tilde N'})^2=0.191$. For the averaged form it equals $8c_P+2c_Q$, since $\mathrm{Tr}[(\hat P_{ij}-\tfrac13)\hat P_{ij}]=8$ and $\mathrm{Tr}[(\hat P_{ij}-\tfrac13)\sum_m\hat n_{i,m}\hat n_{j,m}]=2$. Hence any sequence realizing the desired model obeys

\begin{equation}
J'\chi=K\equiv(\sigma_\mu^2+3\overline{\mu^2})/8=0.145.
\label{eq:si_K}
\end{equation}
The $\mathfrak{su}(3)$ protection is therefore intrinsic and pulse sequences can only redistribute $K$ between $J'$ and $\chi$.

\emph{Pulse sequence.} %
We illustrate in Fig.~\ref{fig:si_mol}(c) the pulse sequence realizing the desired Hamiltonian. In the basis $\ket{-1},\ket{0},\ket{+1}$, it is built from the single-molecule unitaries
\begin{gather}
X=\begin{pmatrix}0&1&0\\0&0&1\\1&0&0\end{pmatrix},\qquad
W=\frac1{\sqrt3}\begin{pmatrix}1&1&e^{2\pi i/3}\\1&e^{2\pi i/3}&1\\ i&e^{-i\pi/6}&e^{-i\pi/6}\end{pmatrix},\nonumber\\
Z=\begin{pmatrix}-1&0&0\\0&1&0\\0&0&-1\end{pmatrix},\qquad
D_\theta=\begin{pmatrix}e^{-i\theta}&0&0\\0&1&0\\0&0&e^{-i\theta}\end{pmatrix}.
\label{eq:si_ops}
\end{gather}
Here $X$ cyclically shifts the states $\ket{-1}\to\ket{+1}\to\ket{0}\to\ket{-1}$, $Z=e^{i\pi\hat s_z}$ and $D_\theta=e^{-i\theta\hat s_z^2}$ apply diagonal phases, and $W$ maps the $\hat s_z$ basis onto a mutually unbiased one, $|W_{mm'}|^2=\tfrac13$~\cite{Wootters1989}.

The cycle in Fig.~\ref{fig:si_mol}(c) contains two blocks of six equally weighted frames. Block I (weight $w_{\rm I}=\tfrac14$) uses the frames $X^sZ^k$ ($s=0,1,2$; $k=0,1$). Averaging over the cyclic shifts $X^s$ gives all exchange channels the same amplitude and turns the Ising interaction into the density--density term $\sum_m\hat n_{i,m}\hat n_{j,m}$. Averaging over $Z^k$ removes the additional, undesired pair processes into which $X$ maps the pair term. %
Block II (weight $w_{\rm II}=\tfrac34$) uses the frames $WX^sZ^kD_\theta$. It repeats the average of block I after the basis change $W$, and $D_\theta$ multiplies its pair term $\hat\Pi^+_{ij}$ by $e^{2i\theta}$ while leaving $\hat P_{ij}$ and the density-density term unchanged.

The block averages are
\begin{align}
a_{\mathrm{I}}&=J, & a_{\mathrm{II}}&=e^{2i\theta-i\pi/6}\big[J+\tfrac i3(\overline{\mu^2}-\sigma_\mu^2)\big],\nonumber\\
c_P^{\mathrm{I}}&=\overline{\mu^2}, & c_P^{\mathrm{II}}&=\tfrac13(\sigma_\mu^2+2\overline{\mu^2}),\nonumber\\
c_Q^{\mathrm{I}}&=\sigma_\mu^2-\overline{\mu^2}, & c_Q^{\mathrm{II}}&=-\tfrac13(\sigma_\mu^2-\overline{\mu^2}),
\end{align}
with $J=\mu_{01}\mu_{12}/3=0.0590$. Critically, $c_Q^{\mathrm{II}}=-c_Q^{\mathrm{I}}/3$ for any real $\mu$, so the weights $w_{\mathrm{II}}=3w_{\mathrm{I}}$ cancel the undesired density-density term exactly. Choosing $\theta=\tfrac{\pi}{12}-\tfrac12\arg[J+\tfrac i3(\overline{\mu^2}-\sigma_\mu^2)]$ makes $a_{\mathrm{II}}$ real and positive, so the pair amplitudes of the two blocks add, $J'=w_{\mathrm{I}}a_{\mathrm{I}}+w_{\mathrm{II}}a_{\mathrm{II}}$.

Thus, we obtain the gap-protected model with $V=J'E_{\rm dd}$, $\chi=K/J'=2.36$ and
\begin{equation}
J'=\frac{|J|}4+\frac34\sqrt{J^2+\Big(\frac{\overline{\mu^2}-\sigma_\mu^2}3\Big)^2}=0.0615.
\end{equation}

Finally, the ratio of gap-protection strength to pair-creation coupling can be tuned by running two versions of the cycle of Fig.~\ref{fig:si_mol}(c) in sequence. %
In the first cycle we perform the original sequence for a total time multiplied by $1-\eta$. %
For the second cycle, we first apply $D_{\pi/2}=\mathrm{diag}(-i,1,-i)$, then run the original cycle with total time multiplied by $\eta$, and end by applying $D_{\pi/2}^\dagger$. %
In the second cycle every frame $U_f$ therefore becomes $U_fD_{\pi/2}$, i.e., $X^sZ^kD_{\pi/2}$ in block I and $WX^sZ^kD_{\theta+\pi/2}$ in block II. Since $D_{\pi/2}$ multiplies $\hat\Pi^+_{ij}$ by $e^{i\pi}=-1$ and leaves $\hat P_{ij}$ and the density-density term unchanged, the second cycle contributes the pair term with the opposite sign,
\begin{align}
h^{\mathrm{eff}}_{ij}&\to(1-\eta)\,h^{\mathrm{eff}}_{ij}+\eta\,D_{\pi/2}^{(ij)\dagger}h^{\mathrm{eff}}_{ij}D_{\pi/2}^{(ij)}\nonumber\\
&=(1-2\eta)J'\hat\Pi_{ij}+\chi J'\big(\hat P_{ij}-\tfrac13\big)+\mathrm{const}.
\end{align}
As a consequence, $V\to(1-2\eta)J'E_{\mathrm{dd}}$ and $\chi\to\chi/(1-2\eta)$, allowing access to any value $\chi\ge2.36$ for $0\le\eta<\tfrac12$.

\section{Operator Definitions}\label{app:operators}
In this appendix we provide explicit expressions for the operators used in this work.

\subsection{Gell-Mann basis}
The local operator space of a spin-1 particle is spanned by the 8 Gell-Mann matrices $\lambda_a$. These are traceless, Hermitian $3 \times 3$ matrices that satisfy the orthogonality condition $\text{Tr}(\lambda_a \lambda_b) = 2 \delta_{ab}$. The collective operators for $N$ particles are $\Lambda_a = \sum_{k=1}^N \lambda_a^{(k)}$.  

In the standard $s_z$ basis $\{|+1\rangle, |0\rangle, |-1\rangle\}$, the matrices are explicitly defined as:
\begin{align}
\lambda_1 &= 
\begin{pmatrix} 
0 & 1 & 0 \\ 
1 & 0 & 0 \\ 
0 & 0 & 0 
\end{pmatrix}, &
\lambda_2 &= 
\begin{pmatrix} 
0 & -i & 0 \\ 
i & 0 & 0 \\ 
0 & 0 & 0 
\end{pmatrix}, \nonumber \\
\lambda_3 &= 
\begin{pmatrix} 
1 & 0 & 0 \\ 
0 & -1 & 0 \\ 
0 & 0 & 0 
\end{pmatrix}, &
\lambda_4 &= 
\begin{pmatrix} 
0 & 0 & 1 \\ 
0 & 0 & 0 \\ 
1 & 0 & 0 
\end{pmatrix}, \nonumber \\
\lambda_5 &= 
\begin{pmatrix} 
0 & 0 & -i \\ 
0 & 0 & 0 \\ 
i & 0 & 0 
\end{pmatrix}, &
\lambda_6 &= 
\begin{pmatrix} 
0 & 0 & 0 \\ 
0 & 0 & 1 \\ 
0 & 1 & 0 
\end{pmatrix}, \nonumber \\
\lambda_7 &= 
\begin{pmatrix} 
0 & 0 & 0 \\ 
0 & 0 & -i \\ 
0 & i & 0 
\end{pmatrix}, &
\lambda_8 &= \frac{1}{\sqrt{3}}
\begin{pmatrix} 
1 & 0 & 0 \\ 
0 & 1 & 0 \\ 
0 & 0 & -2 
\end{pmatrix}.
\end{align}

\subsubsection{Pair Hamiltonian in terms of Gell-Mann Operators}

We may equivalently express the Hamiltonian Eq.~\eqref{eq:H_T} in this basis as
\begin{equation}
    H = \frac{1}{2} \sum_{i \neq j} V_{ij} \left( \hat{\lambda}_1^{(i)} \hat{\lambda}_6^{(j)} + \hat{\lambda}_2^{(i)} \hat{\lambda}_7^{(j)} \right),    \label{eq:H_power-law}
\end{equation}

\subsection{Spin and Quadrupolar Operators}
We may alternatively use the basis of spin and quadrupolar operators. In our conventions the spin operators read
\begin{align}
\hat s_x&=\frac{1}{\sqrt2}\begin{pmatrix}0&1&0\\1&0&1\\0&1&0\end{pmatrix},\; \quad
\hat s_y=\frac{1}{\sqrt2}\begin{pmatrix}0&-i&0\\i&0&-i\\0&i&0\end{pmatrix},\; \\
\hat s_z & =\begin{pmatrix}1&0&0\\0&0&0\\0&0&-1\end{pmatrix}.
\end{align}

The local nematic (quadrupole) operators are $q_{i,\mu\nu} = s_{i,\mu} s_{i,\nu} + s_{i,\nu} s_{i,\mu} - \frac{4}{3}\delta_{\mu \nu} \mathbf{1}_i$ with $\mu,\nu\in\{x,y,z\}$, and the collective spin (dipole) and nematic operators are
\begin{equation}
    S_{\mu} = \sum_{i=1}^N s_{i,\mu}, \quad Q_{\mu \nu} = \sum_{i=1}^N q_{i,\mu\nu}.
\end{equation}

For use below we note the following expansions of spin and quadrupolar operators in Gell-Mann matrices
\begin{align}
    S_x &= \frac{1}{\sqrt{2}}(\Lambda_1 + \Lambda_6), & Q_{xz} &= \frac{1}{\sqrt{2}}(\Lambda_1 - \Lambda_6), \label{eq:supp_Sx_Qxz}\\
    S_y &= \frac{1}{\sqrt{2}}(\Lambda_2 + \Lambda_7), & Q_{yz} &= \frac{1}{\sqrt{2}}(\Lambda_2 - \Lambda_7). \label{eq:supp_Sy_Qyz}
\end{align}

\section{Dynamics in terms of spin and phase space quadratures}\label{app:quadratures}
In this appendix we provide a derivation of the mapping of the original Hamiltonian to spin operators and to phase-space quadratures, as well as the resulting dynamics.

\subsection{Hamiltonian in terms of spin-nematic operators}
In the infinite-range limit, Eq.~\eqref{eq:H_power-law} becomes $H_{\text{coll}} = \frac{V}{2}\sum_{i\neq j}(\lambda_1^{(i)}\lambda_6^{(j)}+\lambda_2^{(i)}\lambda_7^{(j)}) = \frac{V}{4}(\Lambda_1\Lambda_6 + \Lambda_6\Lambda_1 + \Lambda_2\Lambda_7 + \Lambda_7\Lambda_2)$, where the on-site terms cancel because $\{\lambda_1,\lambda_6\}=\lambda_4=-\{\lambda_2,\lambda_7\}$. To recast this in spin-nematic operators, we note
\begin{align}
    S_x^2 &= \frac{1}{2}(\Lambda_1^2 + \Lambda_6^2 + \Lambda_1\Lambda_6 + \Lambda_6\Lambda_1), \\
    Q_{xz}^2 &= \frac{1}{2}(\Lambda_1^2 + \Lambda_6^2 - \Lambda_1\Lambda_6 - \Lambda_6\Lambda_1).
\end{align}
Subtracting these yields $S_x^2 - Q_{xz}^2 = \Lambda_1\Lambda_6 + \Lambda_6\Lambda_1$, and similarly $S_y^2 - Q_{yz}^2 = \Lambda_2\Lambda_7 + \Lambda_7\Lambda_2$. Thus $H_{\rm coll} = \frac{V}{4}(S_x^2+S_y^2-Q_{xz}^2-Q_{yz}^2)$, which we regroup according to the conjugate pairs $(S_x,Q_{yz})$ and $(Q_{xz},S_y)$ identified below,
\begin{equation}
    H_{\text{coll}} = \frac{V}{4} \left[ (S_x^2 - Q_{yz}^2) - (Q_{xz}^2 - S_y^2) \right].
\end{equation}

Because the polar state lacks mean-field magnetization ($\langle \mathbf{S} \rangle = 0$), standard SU(2) squeezing is absent. However, the macroscopic nematic fluctuations defined by the commutators $\langle[S_x, Q_{yz}]\rangle = -2iN$ and $\langle[Q_{xz}, S_y]\rangle = -2iN$ approximately define two independent pairs of canonical conjugate phase-space variables (see below). The collective Hamiltonian thus operates as two independent parametric amplifiers ($H_{\text{coll}} \propto X^2 - P^2$), exponentially squeezing the quantum noise along the diagonals of the respective conjugate phase spaces. 

Defining the squeezed ($\mathcal{S}_-$) and anti-squeezed ($\mathcal{S}_+$) quadratures for the two independent modes, $A$ and $B$, they map to macroscopic spin operators as 
\begin{align}
    \mathcal{S}_\pm^A &= \frac{1}{\sqrt{2N}}(S_x \pm Q_{yz}), \label{eq:quad_A} \\
    \mathcal{S}_\pm^B &= \frac{1}{\sqrt{2N}}(Q_{xz} \mp S_y). \label{eq:quad_B}
\end{align}
In the linearized (undepleted-pump) approximation these exhibit exponentially decreasing ($-$) and increasing ($+$) variances, $\text{Var}[\mathcal{S}_{\pm}] = e^{\pm 2\kappa t}$, with collective rate $\kappa = N V$. Squeezing of $\mathcal{S}_-$ results in Heisenberg-limited metrological gain, $\xi^2 \sim 1/N$, for measurements along these axes once depletion sets in at $e^{2\kappa t}\sim N$.

\subsection{Continuous-Variable Phase Space Mapping}
In the polar state $\vert \psi_0 \rangle = \vert 0 \rangle^{\otimes N}$, expectation values of standard commutators like $[S_x, S_y] = iS_z$ vanish since $\langle S_z \rangle = 0$. For the spin and nematic commutators we have
\begin{align}
    [Q_{xz}, S_y]  &= -2i \sum_{k=1}^N (s_x^2 - s_z^2)^{(k)}, \\
    [S_{x}, Q_{yz}] &= -2i \sum_{k=1}^N (s_y^2 - s_z^2)^{(k)}.
\end{align}

Evaluating this in the polar state $\vert 0 \rangle^{\otimes N}$, where the single-particle expectations are $\langle s_x^2 \rangle = \langle s_y^2 \rangle = 1$ and $\langle s_z^2 \rangle = 0$, results in 
\begin{align}
    \langle[S_x, Q_{yz}]\rangle &= -2iN, \label{eq:supp_comm1} \\
    \langle[Q_{xz}, S_y]\rangle &= -2iN. \label{eq:supp_comm2}
\end{align}

Motivated by Eqs.~\eqref{eq:supp_comm1} and \eqref{eq:supp_comm2}, we define properly normalized phase-space quadratures $[X, P] = i$ for the two independent modes
\begin{align}
    X_A &= \frac{S_x}{\sqrt{2N}}, \quad P_A = \frac{-Q_{yz}}{\sqrt{2N}}, \label{eq:supp_XP1} \\
    X_B &= \frac{Q_{xz}}{\sqrt{2N}}, \quad P_B = \frac{-S_y}{\sqrt{2N}}. \label{eq:supp_XP2}
\end{align}
Substituting these into the Hamiltonian recasts it as the sum of two independent parametric amplifiers
\begin{equation}
    H_{\text{coll}} =\frac{\kappa}{2}  (X_A^2 - P_A^2) - \frac{\kappa}{2} (X_B^2 - P_B^2),
\end{equation}
where $\kappa =V N$ is the effective collective coupling. 

\subsection{Squeezing Dynamics}
Using the Heisenberg equation of motion $\dot{\mathcal{O}} = i[H_{\text{coll}}, \mathcal{O}]$, the eigenmodes are straightforwardly found. For mode A, $\frac{d}{dt}(X_A \pm P_A) = \mp \kappa (X_A \pm P_A)$. This yields the squeezed ($-$) and anti-squeezed ($+$) solutions
\begin{align}
    X_A(t) + P_A(t) &= (X_A(0) + P_A(0)) e^{-\kappa t}, \\
    X_A(t) - P_A(t) &= (X_A(0) - P_A(0)) e^{+\kappa t}.
\end{align}
Mapping back to spin operators, the squeezed ($\mathcal{S}_-$) and anti-squeezed ($\mathcal{S}_+$) quadratures correspond to 
\begin{align}
    \mathcal{S}_-^A &= X_A + P_A = \frac{1}{\sqrt{2N}} (S_x - Q_{yz}), \\
    \mathcal{S}_+^A &=X_A - P_A = \frac{1}{\sqrt{2N}} (S_x + Q_{yz}).
\end{align}
For mode B the relative minus sign in the Hamiltonian interchanges the roles of the two combinations, $\frac{d}{dt}(X_B \mp P_B) = \mp \kappa (X_B \mp P_B)$, so that
\begin{align}
    \mathcal{S}_-^B &= X_B - P_B = \frac{1}{\sqrt{2N}} (Q_{xz} + S_y), \\
    \mathcal{S}_+^B &= X_B + P_B = \frac{1}{\sqrt{2N}} (Q_{xz} - S_y),
\end{align}
reproducing Eq.~\eqref{eq:quad_B}. The variances evolve as $\mathrm{Var}[\mathcal{S}_\pm^{A,B}](t) = e^{\pm 2\kappa t}$, starting from the polar-state value $1$.

\section{Generalized Optimal Squeezing Parameter}\label{app:squeezing}

In this section we define the generalized squeezing parameter optimized over linear generator and read-out operators.

We follow the methodology of Ref.~\cite{Gessner2019}, taking as our accessible set the collective $\mathfrak{su}(3)$ operators of $N$ spin-1 particles. These operators are normalized such that $\mathrm{tr}(\hat\lambda_a\hat\lambda_b)=2\delta_{ab}$, and obey the same algebra as their single-particle counterparts.
\begin{equation}
  \hat\Lambda_a = \sum_{i=1}^{N}\hat\lambda_a^{(i)},
  \qquad
  [\hat\Lambda_a,\hat\Lambda_b] = 2i f_{abc}\hat\Lambda_c ,
  \qquad a = 1,\dots,8 .
  \label{eq:collective}
\end{equation}

We consider the following metrological protocol. A phase is imprinted onto the state by a linear generator $\hat H_{\mathbf n}=\mathbf n\cdot\hat{\boldsymbol\Lambda}$ and subsequently read out from a linear observable $\hat M_{\mathbf m}=\mathbf m\cdot\hat{\boldsymbol\Lambda}$.
The sensitivity of this protocol is then given by
\begin{equation}
  (\Delta\theta)^{-2}(\mathbf m,\mathbf n)
  =\frac{\big|\langle[\hat M_{\mathbf m},\hat H_{\mathbf n}]\rangle\big|^2}
         {(\Delta \hat M_{\mathbf m})^2}
  =\frac{\big(\mathbf m^{T}\mathbf C\,\mathbf n\big)^2}
         {\mathbf m^{T}\boldsymbol\Gamma\,\mathbf m} ,
  \label{eq:sensitivity}
\end{equation}
where we defined
\begin{align}
  \Gamma_{ab} &= \tfrac12\big\langle\{\hat\Lambda_a,\hat\Lambda_b\}\big\rangle
                 - \langle\hat\Lambda_a\rangle\langle\hat\Lambda_b\rangle ,
  \label{eq:Gamma}\\[2pt]
  C_{ab} &= -i\big\langle[\hat\Lambda_a,\hat\Lambda_b]\big\rangle
          = 2 f_{abc}\langle\hat\Lambda_c\rangle .
  \label{eq:C}
\end{align}

Maximizing the sensitivity in Eq.~\eqref{eq:sensitivity} over the read-out direction ($\mathbf{m}$) at a fixed generator has a well-known closed-form solution \cite{Gessner2017,Gessner2019}, $\mathbf m_{\rm opt}\propto\boldsymbol\Gamma^{-1}\mathbf C\,\mathbf n$. This results in 
\begin{equation}
  (\Delta\theta)^{-2}_{\rm opt}(\mathbf n)
  =\max_{\mathbf m}(\Delta\theta)^{-2}(\mathbf m,\mathbf n)
  =\mathbf n^{T}\mathbf M\,\mathbf n   
  \label{eq:M}
\end{equation}
with $\mathbf M = \mathbf C^{T}\boldsymbol\Gamma^{-1}\mathbf C $.

To quantify the metrological gain, this sensitivity must be normalized against the best possible separable state sensing the same generator \cite{Gessner2019}. For a linear generator, this bound is directly determined  by the spectral width of the single-particle generator $h_{\mathbf n}=\mathbf n\cdot\hat{\boldsymbol\lambda}$. Thus,
\begin{equation}
  F_{\rm SN}[\hat H_{\mathbf n}]
  =\max_{\hat\rho_{\rm cl}} F_Q[\hat\rho_{\rm cl},\hat H_{\mathbf n}]
  = N\,\delta h(\mathbf n)^2 ,
\end{equation}
with $\delta h(\mathbf n)=\lambda_{\max}(h_{\mathbf n})-\lambda_{\min}(h_{\mathbf n})$.

Therefore the optimal squeezing parameter measuring the metrological gain takes the form
\begin{equation}
  \xi^{2}_{\rm opt}
  =\min_{\mathbf n}
  \frac{F_{\rm SN}[\hat H_{\mathbf n}]}{(\Delta\theta)^{-2}_{\rm opt}(\mathbf n)}
  =\min_{\mathbf n}
  \frac{N\,\delta h(\mathbf n)^2}{\mathbf n^{T}\mathbf M\,\mathbf n}
  \label{eq:xi2}
\end{equation}

For spin-1/2 particles Eq.~\eqref{eq:xi2} reduces to the familiar Wineland squeezing parameter \cite{Wineland1992}, since then $\delta h(\mathbf n)=2$ for all unit vectors $\mathbf n$ and the optimization reduces to choosing $\mathbf n$ along the eigenvector of $\mathbf M$ with the largest eigenvalue. For the full set of $\mathfrak{su}(3)$ generators, $\delta h(\mathbf n)$ varies between $\sqrt{3}$ and $2$, and in general the optimal $\vect{n}$ is not an eigenvector of $\mathbf{M}$, and the ratio has to be optimized numerically over $\mathbf{n}$.

\section{GDTWA Benchmarks} \label{app:gdtwa}

Since GDTWA is an approximate method, we benchmark its performance for the Hamiltonian in Eq.~\eqref{eq:H_power-law}. Specifically, we consider one-dimensional systems with interaction exponents $\alpha = 0, 1, 2, 3$, and compare GDTWA results against exact calculations for both the expectation values $\langle \hat{\Lambda}_{a} \rangle$ and the symmetrized correlations $\tfrac12\langle \{\hat{\Lambda}_{a},\hat{\Lambda}_{b}\} \rangle$ entering $\Gamma_{ab}$. 

\begin{figure}[tbph]
    \centering
    \includegraphics[width=0.9\linewidth]{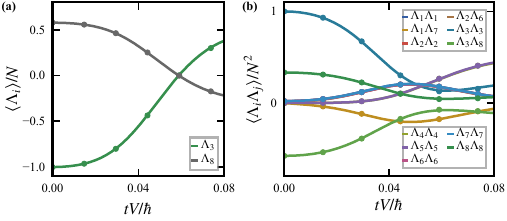}
 \caption{Benchmark: GDTWA (lines) versus exact results (markers) for $\alpha=0$, $N=50$. (a) Time evolution of collective observables $\langle\Lambda_a\rangle$ and (b) symmetrized correlations $\langle \{\Lambda_a, \Lambda_b \}/2\rangle$.
    \label{fig:Supplementary_benchmarking_alpha0}}
\end{figure}

We first consider the all-to-all interaction limit ($\alpha = 0$). For a system of $N=50$ particles in 1D, we find that the GDTWA results are in excellent agreement with the exact calculations, with the expectation values $\langle \hat{\Lambda}_{\alpha} \rangle$ and correlations $\langle \hat{\Lambda}_{\alpha\beta} \rangle$ overlapping throughout the relevant interaction timescales, as shown in Fig.~\ref{fig:Supplementary_benchmarking_alpha0}. Such agreement is expected since all-to-all interactions give rise to collective dynamics, a regime that is particularly well captured by GDTWA.

\begin{figure}[tbph]
    \centering
    \includegraphics[width=0.9\linewidth]{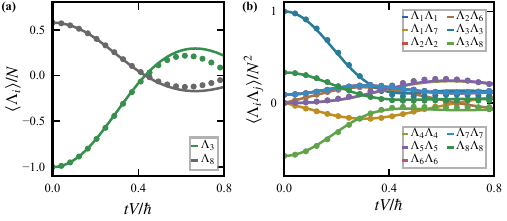}
\caption{Benchmark: GDTWA (lines) versus exact results (markers) for $\alpha=1$, $N=11$.  (a) Time evolution of collective observables $\langle\Lambda_a\rangle$ and (b) symmetrized correlations $\langle \{\Lambda_a, \Lambda_b \}/2\rangle$.}
\label{fig:Supplementary_benchmarking_alpha1}
\end{figure}

\begin{figure}[tbph]
    \centering
    \includegraphics[width=0.9\linewidth]{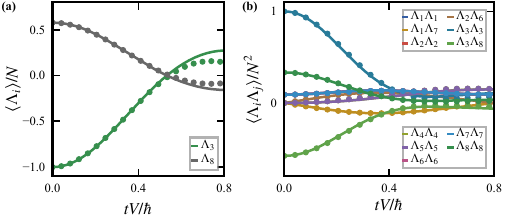}
\caption{Benchmark: GDTWA (lines) versus exact results (markers) for $\alpha=2$, $N=11$.  (a) Time evolution of collective observables $\langle\Lambda_a\rangle$ and (b) symmetrized correlations $\langle \{\Lambda_a, \Lambda_b \}/2\rangle$.}
\label{fig:Supplementary_benchmarking_alpha2}
\end{figure}

\begin{figure}[tbph]
    \centering
    \includegraphics[width=0.9\linewidth]{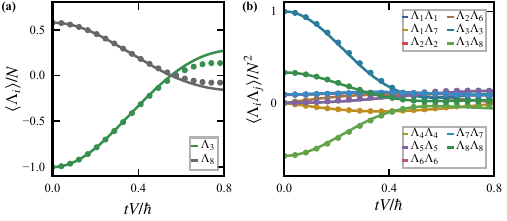}
\caption{Benchmark: GDTWA (lines) versus exact results (markers) for $\alpha=3$, $N=11$.  (a) Time evolution of collective observables $\langle\Lambda_a\rangle$ and (b) symmetrized correlations $\langle \{\Lambda_a, \Lambda_b \}/2\rangle$.}
\label{fig:Supplementary_benchmarking_alpha3}
\end{figure}

For finite-range interactions ($\alpha = 0,1, 2,3$), we compare GDTWA and exact diagonalization for $N=11$ spins. We observe good agreement at short and intermediate times relevant for the optimal squeezing. Deviations begin to emerge at later times not relevant for the squeezing dynamics we are interested in here. Only non-zero expectation values are shown in Figs.~\ref{fig:Supplementary_benchmarking_alpha1}-~\ref{fig:Supplementary_benchmarking_alpha3}.

\begin{figure}[tbph]
     \centering     \includegraphics[width=0.9\linewidth]{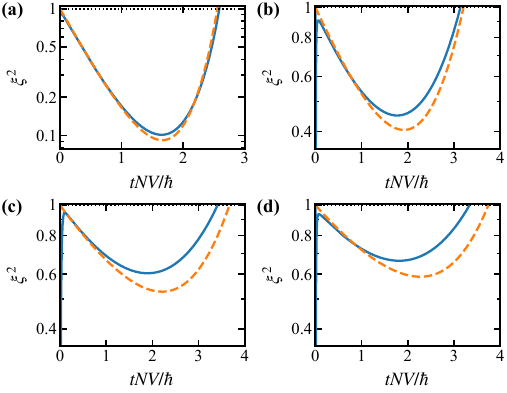}
     \caption{Benchmark of squeezing dynamics. GDTWA (lines) versus exact results (dashed). Time evolution of squeezing parameter $\xi^2$ for all-to-all ($\alpha=0$) interactions in (a) for $L=50$ and for power-law interactions with $\alpha = 1, 2, 3$ in (b)-(d) for $L=11$.}
     \label{fig:SFigS5}
\end{figure}

We also benchmark the GDTWA results against exact dynamics by comparing the time evolution of the squeezing parameter $\xi^2$ shown in Fig.~\ref{fig:SFigS5}. For the all-to-all interacting case, $\alpha=0$ [panel (a)], GDTWA closely reproduces the exact dynamics, including both the optimal squeezing time as well as the squeezing minimum. For finite-range interactions, $\alpha=1,2,$ and $3$ [panels (b)-(d)], GDTWA continues to capture the overall squeezing dynamics and the location of the minimum, although quantitative deviations from the exact result become more apparent near the point of optimal squeezing for small L. Nevertheless, the agreement demonstrates that GDTWA captures the qualitative behavior of the squeezing dynamics across the interaction regimes considered here, and is most quantitatively accurate in the collective regime.

Overall, these benchmarks demonstrate that GDTWA provides a reliable description of the system dynamics over the timescales relevant to our study, particularly for the larger system sizes and more collective regimes that are the primary focus of this work.

\section{Bogoliubov Theory: Scaling of the critical gap-protection strength}
\label{app:critical-chi}

The Bogoliubov spectrum in the main text gives the critical protection
strength
\begin{equation}
    \chi_c=\max_{k \neq 0} \frac{|V_k|}{V_0 - V_k},
    \label{eq:supp-chic-definition}
\end{equation}
above which all finite-momentum modes are stable.  Here we determine the
system-size dependence of $\chi_c$ for
power-law couplings $V(r)=r^{-\alpha}$ on a $d$-dimensional lattice of
side length $L$.  Distances and interaction energies are expressed in the
same dimensionless units as in the main text.  Thus,
\begin{equation}
    V_k=\sum_{\boldsymbol{r}\neq 0}
    \frac{e^{-i\boldsymbol{k}\cdot\boldsymbol{r}}}{r^\alpha},
    \qquad V_0\equiv V_{k=0}.
    \label{eq:supp-vk-lattice}
\end{equation}
For the couplings considered here, $V_k$ decreases away from $k=0$, and
the maximum
in Eq.~\eqref{eq:supp-chic-definition} occurs at the smallest nonzero
momentum, $k_1=2\pi/L$.  The scaling of $\chi_c$ is therefore fixed by the
large-$L$ behavior of $V_0$ and $V_0-V_{k_1}$.

In the continuum approximation, the lattice spacing $a$ and the system size
$L$ provide ultraviolet and infrared cutoffs, respectively:
\begin{equation}
    V_k\simeq \Omega_d\int_a^L dr\,r^{d-1-\alpha}
    \left\langle e^{-i\boldsymbol{k}\cdot\boldsymbol{r}}\right\rangle_\Omega,
    \qquad
    \Omega_d=\frac{2\pi^{d/2}}{\Gamma(d/2)}.
    \label{eq:supp-vk-continuum}
\end{equation}
The angular average can be written as
\begin{equation}
    \left\langle e^{-i\boldsymbol{k}\cdot\boldsymbol{r}}\right\rangle_\Omega
    =\Gamma(d/2)\left(\frac{kr}{2}\right)^{1-d/2}J_{d/2-1}(kr),
    \label{eq:supp-angular-average}
\end{equation}
whose limiting behavior is
\begin{equation}
    \left\langle e^{-i\boldsymbol{k}\cdot\boldsymbol{r}}\right\rangle_\Omega
    \simeq
    \begin{cases}
        1-(kr)^2/(2d)+\mathcal{O}\!\left((kr)^4\right), & kr\ll 1,\\
        \mathcal{O}\!\left((kr)^{-(d-1)/2}\right), & kr\gg 1,
    \end{cases}
    \label{eq:supp-angular-limits}
\end{equation}
with an oscillatory coefficient in the second line.  The uniform component is
therefore
\begin{equation}
    \begin{aligned}
    V_0&=\Omega_d\int_a^L dr\,r^{d-1-\alpha}\\
    &=
    \begin{cases}
        \dfrac{\Omega_d}{d-\alpha}
        \left(L^{d-\alpha}-a^{d-\alpha}\right), & \alpha\neq d,\\[4pt]
        \Omega_d\ln(L/a), & \alpha=d.
    \end{cases}
    \end{aligned}
    \label{eq:supp-v0}
\end{equation}
and hence
\begin{equation}
    V_0\sim
    \begin{cases}
        L^{d-\alpha}, & \alpha<d,\\
        \ln L, & \alpha=d,\\
        \text{const.}, & \alpha>d.
    \end{cases}
    \label{eq:supp-v0-scaling}
\end{equation}

To obtain the dispersion near $k=0$, we split the radial integral for
$V_0-V_k$ at $r=1/k$.  Using Eq.~\eqref{eq:supp-angular-limits} on either
side gives, at the level of scaling,
\begin{equation}
    V_0-V_k\simeq
    \underbrace{\frac{\Omega_d k^2}{2d}
        \int_a^{1/k}dr\,r^{d+1-\alpha}}_{I_{\mathrm{UV}}}
    +
    \underbrace{\Omega_d
        \int_{1/k}^{L}dr\,r^{d-1-\alpha}}_{I_{\mathrm{IR}}}.
    \label{eq:supp-uv-ir-split}
\end{equation}
The two contributions obey
\begin{align}
    I_{\mathrm{UV}}&\sim
    \begin{cases}
        k^{\alpha-d}, & \alpha<d+2,\\
        k^2\ln(1/ka), & \alpha=d+2,\\
        k^2, & \alpha>d+2,
    \end{cases}
    \label{eq:supp-iuv-scaling}\\
    I_{\mathrm{IR}}&\sim
    \begin{cases}
        L^{d-\alpha}, & \alpha<d,\\
        \ln(kL), & \alpha=d,\\
        k^{\alpha-d}, & \alpha>d.
    \end{cases}
    \label{eq:supp-iir-scaling}
\end{align}
At the softest finite momentum, $k_1=2\pi/L$, these results combine to
\begin{equation}
    V_0-V_{k_1}\sim
    \begin{cases}
        L^{d-\alpha}, & \alpha<d+2,\quad \alpha\neq d,\\
        \text{const.}, & \alpha=d,\\
        L^{-2}\ln L, & \alpha=d+2,\\
        L^{-2}, & \alpha>d+2.
    \end{cases}
    \label{eq:supp-delta-v-scaling}
\end{equation}
The logarithms at $\alpha=d$ and $\alpha=d+2$ mark, respectively, the
infrared and ultraviolet marginal cases.

Using $V_{k_1}=V_0-(V_0-V_{k_1})$ in
Eq.~\eqref{eq:supp-chic-definition}, we obtain
\begin{equation}
    \chi_c\sim
    \begin{cases}
        \text{const.}, & \alpha<d,\\
        \ln L, & \alpha=d,\\
        L^{\alpha-d}, & d<\alpha<d+2,\\
        L^2/\ln L, & \alpha=d+2,\\
        L^2, & \alpha>d+2.
    \end{cases}
    \label{eq:supp-chic-scaling}
\end{equation}
For $\alpha>d$, $V_{k_1}$ approaches the nonzero constant $V_0$, whereas
$V_0-V_{k_1}$ vanishes as $L$ increases.  The protection required to
stabilize the softest finite-momentum mode must therefore grow with system
size.  For $\alpha<d$, by contrast, both quantities are dominated by the
far field and scale as $L^{d-\alpha}$.  Their ratio tends to a finite
constant, so a size-independent protection strength is sufficient, although
nonzero protection is still required.  For example, the lattice sums give
$V_{k_1}/V_0\to 0.374$ for $d=1$ and $\alpha=0.5$.

The leading finite-size correction in the long-range regime follows by
writing
\begin{equation}
    V_0=a_0L^{d-\alpha}+a_1+\cdots,
    \qquad
    V_0-V_{k_1}=b_0L^{d-\alpha}+b_1+\cdots.
    \label{eq:supp-subleading-expansions}
\end{equation}
Substitution into Eq.~\eqref{eq:supp-chic-definition} gives
\begin{equation}
    \begin{aligned}
        \chi_c(L)&=\chi_\infty+cL^{-(d-\alpha)}+\cdots,
        \qquad \alpha<d,\\
        \chi_\infty&=\frac{a_0-b_0}{b_0},
    \end{aligned}
    \label{eq:supp-chic-finite-size}
\end{equation}
where $c$ is nonuniversal.  Thus the exponent $\alpha-d$, which controls
the divergence for $d<\alpha<d+2$, reappears below $\alpha=d$ as the
finite-size correction to a plateau.

We test Eqs.~\eqref{eq:supp-chic-scaling} and
\eqref{eq:supp-chic-finite-size} by evaluating $V_k$ directly on finite
periodic lattices using the minimum-image convention.  The resulting fits,
summarized in Table~\ref{tab:supp-chic-fits}, agree very well with the
analytical predictions; the modest deviations in three dimensions reflect
the more restricted range of accessible $L$.

\begin{table}[t]
    \caption{Fits of the exact lattice results to the predicted
    asymptotic forms. A dash denotes a logarithmic fit.}
    \label{tab:supp-chic-fits}
    \centering
    \begingroup
    \scriptsize
    \setlength{\tabcolsep}{2.5pt}
    \renewcommand{\arraystretch}{0.95}
    \begin{tabular}{ccclcc}
        \hline\hline
        $d$ & $\alpha$ & Regime & Fit form & $p$ (fit) & $p$ (predicted) \\
        \hline
        1 & 0.50 & $\alpha<d$ & $\chi_\infty+cL^p$, $\chi_\infty=0.598$ & $-0.500$ & $-0.50$ \\
        1 & 0.75 & $\alpha<d$ & $\chi_\infty+cL^p$, $\chi_\infty=1.806$ & $-0.250$ & $-0.25$ \\
        1 & 1.00 & $\alpha=d$ & $A\ln L+B$, $A=0.606$ & -- & -- \\
        1 & 1.25 & $d<\alpha<d+2$ & $AL^p+B$, $A=1.986$ & $0.250$ & $0.25$ \\
        1 & 1.50 & $d<\alpha<d+2$ & $AL^p+B$, $A=0.786$ & $0.500$ & $0.50$ \\
        1 & 2.00 & $d<\alpha<d+2$ & $AL^p+B$, $A=0.216$ & $1.000$ & $1.00$ \\
        1 & 2.50 & $d<\alpha<d+2$ & $AL^p+B$, $A=0.060$ & $1.493$ & $1.50$ \\
        1 & 3.00 & $\alpha=d+2$ & $AL^p/\ln L+B$, $A=0.068$ & $1.994$ & $2.00$ \\
        1 & 4.00 & $\alpha>d+2$ & $AL^p+B$, $A=0.034$ & $2.000$ & $2.00$ \\
        1 & 5.00 & $\alpha>d+2$ & $AL^p+B$, $A=0.044$ & $2.000$ & $2.00$ \\
        \hline
        2 & 1.00 & $\alpha<d$ & $\chi_\infty+cL^p$, $\chi_\infty=0.452$ & $-0.999$ & $-1.00$ \\
        2 & 1.50 & $\alpha<d$ & $\chi_\infty+cL^p$, $\chi_\infty=1.366$ & $-0.492$ & $-0.50$ \\
        2 & 2.00 & $\alpha=d$ & $A\ln L+B$, $A=0.906$ & -- & -- \\
        2 & 2.50 & $d<\alpha<d+2$ & $AL^p+B$, $A=1.082$ & $0.514$ & $0.50$ \\
        2 & 3.00 & $d<\alpha<d+2$ & $AL^p+B$, $A=0.328$ & $1.000$ & $1.00$ \\
        2 & 4.00 & $\alpha=d+2$ & $AL^p/\ln L+B$, $A=0.122$ & $1.975$ & $2.00$ \\
        2 & 5.00 & $\alpha>d+2$ & $AL^p+B$, $A=0.064$ & $1.979$ & $2.00$ \\
        2 & 6.00 & $\alpha>d+2$ & $AL^p+B$, $A=0.078$ & $1.999$ & $2.00$ \\
        \hline
        3 & 2.00 & $\alpha<d$ & $\chi_\infty+cL^p$, $\chi_\infty=0.764$ & $-0.940$ & $-1.00$ \\
        3 & 2.50 & $\alpha<d$ & $\chi_\infty+cL^p$, $\chi_\infty=2.000$ & $-0.428$ & $-0.50$ \\
        3 & 3.00 & $\alpha=d$ & $A\ln L+B$, $A=1.068$ & -- & -- \\
        3 & 4.00 & $d<\alpha<d+2$ & $AL^p+B$, $A=0.398$ & $1.005$ & $1.00$ \\
        3 & 5.00 & $\alpha=d+2$ & $AL^p/\ln L+B$, $A=0.170$ & $1.954$ & $2.00$ \\
        3 & 6.00 & $\alpha>d+2$ & $AL^p+B$, $A=0.102$ & $1.932$ & $2.00$ \\
        \hline\hline
    \end{tabular}

    \endgroup
\end{table}

Equation~\eqref{eq:supp-chic-scaling} separates experimentally distinct
regimes.  For $\alpha>d$, the required protection grows without bound and
eventually as $L^2$ for sufficiently short-range interactions.  For
$\alpha=d$, the growth is only logarithmic, while for $\alpha<d$ a fixed
$\chi>\chi_\infty$ stabilizes the collective mode at arbitrary system size.

\section{Extended Data}\label{app:extended_data}

\subsection{Extended Data for Gap Protected Squeezing/Scaling}
Figure~\ref{fig:fig3} presents the scaling exponent $\nu$ for representative interaction exponents, namely $\alpha=2$ in 1D and $\alpha=3$ in 2D and 3D. Here, we extend this analysis to a broader range of $\alpha$ in each spatial dimension in order to examine the robustness of the transition between non-collective and collective scaling.

\begin{figure}[tbp]
    \centering    \includegraphics[width=\linewidth]{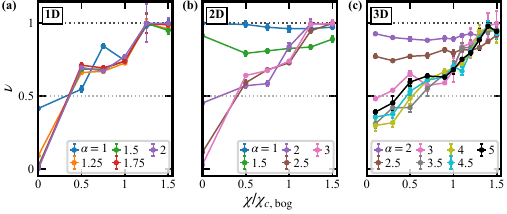}
    \caption{Scaling exponent $\nu$ as a function of the normalized gap-protection strength $\chi/\chi_{c,\mathrm{bog}}$ for varying interaction exponent $\alpha$ in (a) 1D, (b) 2D, and (c) 3D. The horizontal dotted line at $\nu=1$ denotes fully collective scaling. Across all dimensions, increasing the gap-protection strength enhances collective behavior, particularly for larger $\alpha$, where the system exhibits non-collective or partially collective scaling at weak gap protection and approaches fully collective scaling as $\chi$ is increased. Error bars indicate the fitting uncertainty of the extracted scaling exponent $\nu$.} 
    \label{fig:FigS6}
\end{figure}

Figure~\ref{fig:FigS6}(a) shows the 1D results for $\alpha=\{1,1.25,1.5,1.75,2\}$. Across the interaction exponents considered, we find a common transition as a function of $\chi/\chi_{c,\mathrm{bog}}$. For $\chi/\chi_{c,\mathrm{bog}}<1.25$, the extracted exponent satisfies $\nu<1$, indicating sub-collective scaling. In contrast, for $\chi/\chi_{c,\mathrm{bog}}\geq1.25$, $\nu$ approaches unity for all values of $\alpha$, indicating the recovery of collective scaling.

Figure~\ref{fig:FigS6}(b) shows the corresponding results in 2D for $\alpha=\{1,1.5,2,2.5,3\}$. The behavior now depends more strongly on the interaction exponent. For $\alpha<d$, with $d=2$, the system exhibits predominantly collective scaling, with $\nu$ remaining close to unity over the range of $\chi$ considered. In particular, for $\alpha=1$, the scaling remains collective throughout. For $\alpha=1.5$, $\nu$ also remains close to unity, although the finite-size data do not allow us to sharply distinguish small deviations from fully collective scaling. At $\alpha=2$, corresponding to $\alpha=d$, the small-$\chi$ regime exhibits partially collective behavior, with $0<\nu<1$. Increasing the gap-protection strength enhances collectiveness, and $\nu$ approaches unity for $\chi/\chi_{c,\mathrm{bog}}\gtrsim1.25$. For the shorter-ranged cases $\alpha=\{2.5,3.0\}$ the distinction is considerably stronger, at small $\chi$, $\nu$ is close to zero, indicating essentially non-collective scaling. As $\chi$ is increased, $\nu$ increases continuously, demonstrating the restoration of collective behavior by the gap-protection term. For $\chi/\chi_{c,\mathrm{bog}}\gtrsim1.25$, $\nu$ again approaches unity, indicating fully collective scaling.

Since the transition occurs between the last sub-collective point at $\chi/\chi_{c,\mathrm{bog}}=1$ and the first collective point at $\chi/\chi_{c,\mathrm{bog}}=1.25$, we identify their midpoint, %
$
\frac{\chi_c}{\chi_{c,\mathrm{bog}}}=1.125 $%
, as the transition criterion for the data across the range of $\alpha$ considered for 1D and 2D. 

%%% explain 3D %%%%
Figure~\ref{fig:FigS6}(c) shows the 3D results for $\alpha = \{2,2.5,3,3.5,4,4.5,5\}$. As in 2D, the behavior depends on the interaction exponent. For $\alpha < d$, the scaling is already largely collective at small $\chi$, and $\nu$ moves toward unity as the gap-protection strength is increased. At $\alpha = 3$, corresponding to $\alpha = d$, the small-$\chi$ regime is only partially collective. For the shorter-ranged cases $\alpha = \{3.5,4,4.5,5\}$ the difference is much stronger. At small $\chi$, $\nu$ is far below one, so the scaling is essentially non-collective. Increasing $\chi$ raises $\nu$ continuously, which shows that the gap-protection term restores collective behavior. For $\chi/\chi_{c,\mathrm{bog}} \gtrsim 1.45 \pm 0.05$, $\nu$ is close to unity for all values of $\alpha$, indicating fully collective scaling.

Unlike in 1D and 2D, the transition in 3D does not occur at a single value for all $\alpha$. For the longer-ranged cases ($\alpha \leq 3$) it sits at $\chi_c/\chi_{c,\mathrm{bog}} \approx 1.35$, while for the shorter-ranged cases ($\alpha > 3$) it is found at $\chi_c/\chi_{c,\mathrm{bog}} = 1.45 \pm 0.05$. %Here in 3D, we used the same mid-point method as in 1D and 2D to find the transition point. 
In both cases the transition lies between $\chi/\chi_{c,\mathrm{bog}} = 1.35$ and $1.45$, which is higher than the value $\chi_c/\chi_{c,\mathrm{bog}} = 1.125$ obtained in 1D and 2D.

Taken together, panels (a)--(c) show that increasing $\chi$ restores collective scaling in all three dimensions, even when the power-law interactions alone give only partially collective or non-collective behavior.

\subsection{Extended Site Dilution Data}

\begin{figure}[tbp]
    \centering
    \includegraphics[width=\linewidth]{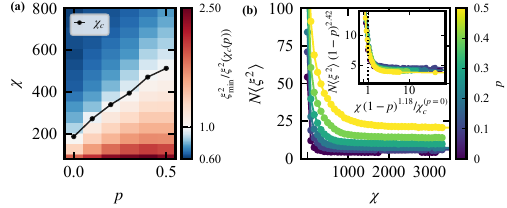}
    \caption{1D disorder spin squeezing phase diagram ($L=100$, $\alpha=3$, 5000 trajectories). (a) Heatmap of the squeezing parameter $\xi^2/ \xi^2(\chi_{c}(p))$ as a function of gap protection strength $\chi$ and dilution probability $p$. Black triangles indicate the disorder-dependent transition $\chi_c(p)$. (b) Squeezing as a function of $\chi$ for different dilution probabilities. The inset shows the corresponding data collapse. The disorder-dependent scaling functions are described by $f(p)=B(1-p)^{-m}$ with  $B=0.04$ and $m=2.42$  for plateau collapse, and $B= 241$ and $m=1.18$ for horizontal collspse.}
    \label{fig:disorder1D}
\end{figure}

We investigate the effect of disorder on the squeezing dynamics in one- and two-dimensional arrays analogous to the 3D case shown in Fig.~\ref{fig:fig4}. Without dilution ($p=0$), the transition from the non-collective to the collective regime is determined from the finite-size scaling analysis. Specifically, from the $\nu$ versus $\chi$ plot, where $\nu$ is the scaling exponent, we identify the first value of $\chi$ at which $\nu$ reaches unity and define the transition point $\chi_{\mathrm{c}}(0)$ as the midpoint between this value and the preceding simulated value of $\chi$. To extend the identification of the transition to finite dilution without performing an independent finite-size scaling analysis for each value of $p$, we use the non-disordered case $p=0$ transition as a reference. We evaluate the squeezing at transition $\chi$ for non-disordered case and quantify its relative deviation from the corresponding plateau squeezing through
\begin{equation}
r=
\frac{\xi^2\!\left(\chi_{\mathrm{c}}(p=0),p=0\right)}
{\xi^2_{\mathrm{plateau}}(p=0)}-1.
\end{equation}
The resulting value of $r$ is $r=0.6$ in 1D and $r=2.4$ in 2D. 

For $p>0$, we then determine the plateau squeezing $\xi^2_{\mathrm{plateau}}(p)$ and identify the disorder-dependent transition $\chi_{\mathrm{c}}(p)$ from the condition
\begin{equation}
\xi^2\!\left(\chi_{\mathrm{c}}(p),p\right)
=(1+r)\,\xi^2_{\mathrm{plateau}}(p).
\end{equation}
Thus, $\chi_{\mathrm{c}}(p)$ corresponds to the value of $\chi$ at which the squeezing reaches the same relative deviation from its respective plateau value as at the clean-system transition, providing a consistent criterion for identifying the transition across different dilution strengths.

\begin{figure}[tb!]
    \centering
    \includegraphics[width=\linewidth]{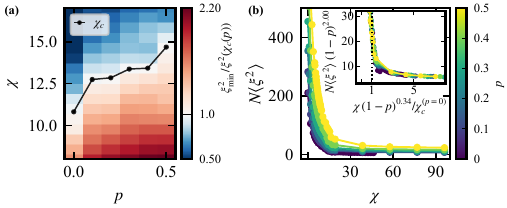}
    \caption{2D disorder spin squeezing phase diagram ($L=32$, $\alpha=3$, 5000 trajectories). (a) Heatmap of the squeezing parameter $\xi^2/ \xi^2(\chi_{c}(p))$ as a function of gap protection strength $\chi$ and dilution probability $p$. Black triangles indicate the disorder-dependent transition $\chi_c(p)$. (b) Squeezing as a function of $\chi$ for different dilution probabilities. The inset shows the corresponding data collapse. The disorder-dependent scales follow $f(p)=B(1-p)^{-m}$ with $B=0.005$ and $m=2.0$ for plateau collapse and, $B=11$ and $m=0.34$ for horizontal rescaling }
    \label{fig:disorder2D}
\end{figure}

The results for the disordered one- and two-dimensional systems are shown in Figs.~\ref{fig:disorder1D} and \ref{fig:disorder2D}, respectively, for $L=100$ in 1D and $L=32$ in 2D, with power-law exponent $\alpha=3$ and dilution probabilities $p=\{0,0.1,0.2,0.3,0.4,0.5\}$. The heat maps in Figs.~\ref{fig:disorder1D}(a) and \ref{fig:disorder2D}(a) show the squeezing as a function of the dilution $p$ and gap-protection strength $\chi$, with black markers indicating the transition values $\chi_c(p)$ for each dilution. In both dimensions, increasing dilution progressively degrades the achievable squeezing and shifts the transition toward larger values of $\chi$, as also evident from the squeezing curves in Figs.~\ref{fig:disorder1D}(b) and \ref{fig:disorder2D}(b). Nevertheless, the systems retain a transition from the non-collective to the collective regime with increasing $\chi$ across all dilutions considered.

To characterize the disorder dependence and compare the squeezing curves across different dilutions, we rescale the horizontal axis by the dilution-dependent transition value and the vertical axis by the corresponding plateau squeezing, $\xi_{\mathrm{opt}}^2( \chi,p)/\xi_{\mathrm{plateau}}^2(p)$. The transition values and plateau squeezing are fitted using the functional form $f(p)=B(1-p)^{-m}$. For the horizontal rescaling, the fitting parameters are $B=241$, $m=1.18$ in 1D and $B=11$, $m=0.34$ in 2D, while for the plateau squeezing they are $B=0.04$, $m=2.42$ in 1D and $B=0.005$, $m=2.0$ in 2D. The horizontal rescaling approximately captures the dilution-dependent shift in the transition, while the vertical rescaling captures the change in the plateau squeezing. Together, these rescalings yield the approximate collapse of the squeezing curves shown in the insets of Figs.~\ref{fig:disorder1D}(b) and \ref{fig:disorder2D}(b), suggesting that the dominant effects of dilution can be captured through the shifts in the transition strength and plateau squeezing over the dilution range considered.

\end{document}